\documentclass[12pt,english]{article}
\usepackage[T1]{fontenc}
\usepackage[latin9]{inputenc}
\usepackage[letterpaper]{geometry}
\usepackage{amsthm}
\usepackage{amsmath}
\usepackage{amsfonts}
\usepackage{amssymb}
\usepackage{graphicx}
\theoremstyle{plain}
\usepackage{babel}
\newcommand{\dd}{\mathrm{d}}

\usepackage{color}

\graphicspath{{../figs/}}

\begin{document}

\begin{titlepage}
\bigskip
\rightline
\bigskip\bigskip\bigskip\bigskip
\centerline {\Large \bf {Charged Randall-Sundrum black holes in Higher Dimensions}}
\bigskip\bigskip
\bigskip\bigskip

\centerline{\large M. Meiers\footnote{mcmeiers@gmail.com}, L. Bovard\footnote{bovard@th.physik.uni-frankfurt.de} and R.B. Mann\footnote{rbmann@uwaterloo.ca}}
\bigskip\bigskip
\centerline{\em Department of Physics \& Astronomy, University of Waterloo, Waterloo, Ontario N2L 3G1, Canada}
\bigskip\bigskip

\begin{abstract}
We extend some solutions for black holes in the Randall-Sundrum theory
with a single brane. We consider a generalised version of the extremal
black hole on the brane in $n+1$ dimensions and determine an asymptotic
value of the geometry for large black holes. 
\end{abstract}

\end{titlepage}

\section{Introduction}
Some time ago Randall and Sundrum proposed  \cite{Randall}  a novel approach to add
an extra unexperienced dimension to use in quantum gravity by considering
a $(4+1)$ dimensional world where non-gravitational physics is
constrained to a $(3+1)$ dimensional hypersurface (or brane), while
gravitational effects are allowed to propagate through a bulk $5$
dimensional AdS spacetime. At low energies, the theory reduces to 4D
general relativity at large distances compared with the AdS length
$\ell$.

In understanding whether
and how the RS model is capable of recovering the strong field
predictions of general relativity for bulk dimensions $N>4$ \cite{RSBH}, 
localized black hole solutions provide a useful tool.
While a considerable amount of numerical work suggests such solutions
exist (at least for small black holes) \cite{numer}\cite{H}, there is still no
analytic solution, except in $(2+1)$ dimensions \cite{EHM}.  However it
is known that there are no asymptotically flat black hole solutions to general relativity in
$(2+1)$ dimensional spacetime, leading to the conclusion that the existence
of this solution is due to quantum corrections from the dual Conformal
Field Theory. These corrections turn what would be a  conical singularity
classically into a regular horizon \cite{EFK}.  While inducing a negative
cosmological constant on the brane yields braneworld black hole solutions
that are similar to those of $(2+1)$ dimensional AdS general relativity
\cite{EHM2},  large black holes are not localized on the brane.
  
In an attempt to make progress on this issue,  Kaus and Reall (KR)
\cite{KausReall}, considered an extreme black hole in (4+1) dimensions
charged with respect to an electromagnetic field on the brane. By
examining the extremal solution one can take advantage of symmetries of
the near horizon geometry in the bulk solution. This approach has the
bulk equations reduce to ordinary differential equations that integrate
to yield a 1-parameter family of solutions.  Solving the Israel junction
conditions to obtain the gravitational effect on the brane yields a
relationship between this parameter and the charge on the brane, which
then serves to label this family of solutions. 

Here we seek to find extremal black hole solutions in $N=(n+1)$ bulk dimensions that contains an electromagnetic field
on an $n$-dimensional brane.  We intend to determine whether the structure exhibited by the $(4+1)$ dimensional extremal black hole is unique to that dimension of the brane and whether pathologies exist in higher dimensional braneworld theories.  There is hope that there is
some robustness in the properties found in models in the style of Randall
and Sundrum, as seen in other higher dimensional extensions of other
models  \cite{AGHKLM} \cite{KS}. For properties that are robust, one 
could hope to better understand them in the large $N$ limit where general
relativity reduces to a simpler model \cite{Emparan2013}. There may also
be advantages of allowing $N$ to take continuous values and look at
limits where the deviation of $N$ from an integer value becomes small. 

We proceed by employing a warped product ansatz for the near-horizon metric whose space transverse to the horizon is a sphere of dimension $D$ (as opposed to 2 in KR) and solve the resulting field equations in this context. By solving the Israel junction conditions on the brane we are able to determine the gravitational affect of the brane on the overall geometry. For $D=2$ the structure of the Israel junction conditions determine the specific geometry of the spacetime, $k=-1$ \cite{KausReall}. These arguments can then be continued to higher dimensions to again restrict to branes of the form $AdS_2\times S^D$.

Using our ansatz we find equations of motion that can be solved with a
single parameter family of solutions. The action on an
electromagnetically charged brane is then used to put junction conditions
on the extrinsic curvature in the bulk. We  then look at the possible
solutions in greater depth and delve into uncovering the structural
relations between the charge and radius of the black hole. We observe how
the entropy measured in the brane and bulk differ, and construct  an
argument to find that for dimensions greater than $5$ the scaling of the
entropy of small black holes is fundamentally different. 

We begin with generalizing the derivations \cite{KausReall} of the
equations of motion and junction conditions for higher dimension.
Following our equations we look  at the  analytic solutions possible
before we consider numeric solutions. We conclude by discussing the
properties of these solutions and examining the entropy scaling observed.

\section{The Equations of Motion}

The near horizon geometry of a static extreme black hole can be written
in the warped product form \cite{KLR}

\begin{equation}\label{NHmet}
ds^{2} = A^{2}(z)  d\Sigma^{2} + dz^{2} + R^{2}(z) d\Omega_{D}^{2}
\end{equation}
with $d\Sigma^{2} = -dt^{2} + S(k)^2dr^{2}$,  where $S(0) = 1$, $S(-1) =
\sin(t)$ and $S(1) = \sinh(t)$ . Both $t$ and $r$ have been made unitless
by using either  the (A)dS radius $\ell$ for $k=\pm1$ or some arbitrary
length scale for $k=0$.  The co-ordinate $z$ corresponds to the
transverse distance from the brane and  $d\Omega_{D}^{2}$ is the line
element on $S^{D}$, with  $N-1=D+2$ the dimension of the brane, and $N =
D+3$  the overall dimension of the bulk. 

The corresponding bulk Einstein field equations are given by

\begin{equation}
R_{\mu\nu} = -\frac{(D+2)}{\ell^{2}}g_{\mu\nu}
\end{equation}

and inserting  the black hole ansatz (\ref{NHmet}) into the field equations yields 

\begin{align}\label{efe}
\frac{k}{A^{2}}- \frac{1}{A^2}\left(\frac{d A}{dz }\right)^2 -  \frac{D}{AR}\frac{d A}{dz }\frac{d R}{dz } -\frac{1}{A}\frac{d^2 A}{dz^2 } &= -\frac{(D+2)}{\ell^{2}}\nonumber\\
\frac{2}{A}\frac{d^2 A}{dz^2 } +\frac{D}{R}\frac{d^2 R}{dz^2 }  &= \frac{(D+2)}{\ell^{2}}\\
\frac{(D-1)}{R^{2}}-\frac{(D-1)}{R^{2}}\left(\frac{d R}{dz }\right)^2-\frac{2}{AR}\frac{d A}{dz }\frac{d R}{dz } -\frac{1}{R}\frac{d^2 R}{dz^2 } &= -\frac{(D+2)}{\ell^{2}}\nonumber
\end{align}

which reduce to the 5-dimensional version of \cite{KausReall} with $D=2$. By denoting $R=\ell\,\rho(z/\ell)$ and $A=\ell\,\alpha(z/\ell)$ we can re-write these equations   as
\begin{align}
\frac{k-\alpha'^{2}}{\alpha^{2}} - \frac{D\alpha'\rho'}{\alpha\rho} &-\frac{\alpha''}{\alpha} = -(D+2)\nonumber\\
\frac{2\alpha''}{\alpha}+\frac{D\rho''}{\rho} &= (D+2)\\
\frac{(D-1)(1-\rho'^{2})}{\rho^{2}}-\frac{2\alpha'\rho'}{\alpha\rho}&-\frac{\rho''}{\rho} = -(D+2)\nonumber
\end{align}
where the prime denotes the deriviative with respect to $x=z/\ell$.
 The Hamiltonian constraint is given by combining these three equations to eliminate the second order derivatives.
\begin{align}\label{hamcon}
\frac{2(k-\alpha'^2)}{\alpha^{2}} -\frac{4D\alpha'\rho'}{\alpha\rho}+ \frac{D(D-1)(1-\rho'^2)}{\rho^{2}} =  - (D+2)(D+1)
\end{align}

Since the horizon is compact in the bulk, the D- sphere of the geometry must contract to a point. As a result $\rho(z/l)$ must vanish somewhere; this location can be chosen to be $z=0$ without loss of generality. Requiring that the equations of motions are smooth at $z=0$ then implies  
\begin{align}\label{series1}
&\alpha(z/\ell) =\sqrt{\frac{\beta}{D+2}}+\frac{\beta+k}{D+1}\left(\frac{D+2}{\beta}\right)^{1/2}\left(\frac{z}{\ell}\right)^2\nonumber \\
&+\frac{\beta+k}{D+1}\left(\frac{(D-1)\beta-(3D+5)k}{(D+1)(D+3)}\right)\left(\frac{D+2}{\beta}\right)^{3/2}\left(\frac{z}{\ell}\right)^4\ldots \\\label{series2}
&\rho(z/\ell)=\frac{z}{\ell}+\frac{(D-1)\beta-2k}{D(D+1)}\left(\frac{D+2}{\beta}\right)\left(\frac{z}{\ell}\right)^3\nonumber\\
&+\frac{\left(D^3+D^2+19D+3\right)\beta^2+4\left(5D^2+16D+3\right)k\beta+4\left(6D^2+13D+3\right)k^2}{D^2(D+1)^2(D+3)}\left(\frac{D+2}{\beta}\right)^2\left(\frac{z}{\ell}\right)^5\ldots
\end{align}
where  $\alpha(0)=\sqrt{\frac{\beta}{D+2}}$ for some positive $\beta$.

\section{Junction Conditions}

The $n=N-1$ dimensional  brane action is given by 
\begin{align}
S_{\text{brane}} = \int d^{n}z\sqrt{-h}\left(-\sigma -\frac{1}{16\pi G_{n}}F_{ij}F^{ij}\right)
\end{align}
where $h_{ij}$ is the induced metric on the brane, $\sigma$ is the brane tension, $G_{n}$ is Newton's constant on the brane, and $F$ is the electromagnetic field on the brane. The tension of the brane is given by $\sigma = (N-2)/4\pi G_{N}\ell$ and $G_n=(N-3)G_N/2\ell$ \cite{EHM2}.
The action results in an energy-momentum tensor 
\begin{align}
& T_{ab}=\frac{1}{4\pi G_n}F_{a i}{F_b}^i-h_{ab}\left(\sigma+\frac{1}{16\pi G_n}F_{ij}F^{ij}\right) \\
& T=-\frac{N-5}{16\pi G_n}F_{ij}F^{ij}-(N-1)\sigma
\end{align}
localized on the brane.
Employing the  junction conditions (see (\ref{eq:IJC}) as in the appendix) and assuming that the brane is located at $z= z_{0}$ is $\mathbb{Z}_2$ symmetric about $z_0$ and thus $K_{ab}(z_0^+)=-K_{ab}(z_0^-)$ we find
\begin{align}\label{junction}
K_{ab}(z_0) &=\frac{8\pi G_N}{2}\left(\frac{1}{N-2}\left(-\frac{N-5}{16\pi G_n}F_{ij}F^{ij}-(N-1)\sigma\right)h_{ab}+\frac{-1}{4\pi G_n}F_{a i}{F_b}^i\right.\nonumber \\ &+\left. h_{ab}\left(\sigma+\frac{1}{16\pi G_n}F_{ij}F^{ij}\right) \right) \nonumber \\
&=\frac{-4\pi G_N \sigma}{N-2}h_{ab}+\frac{G_N}{G_n}\left(\frac{3}{4(N-2)}h_{ab}F_{ij}F^{ij}-F_{a i}{F_b}^i  \right)\nonumber \\
&=\frac{-1}{\ell}h_{ab}+\frac{2\ell}{N-3}\left(\frac{3}{4(N-2)}F_{ij}F^{ij}h_{ab}-F_{a i}{F_b}^i\right)
\end{align}
which up to a sign convention reduces down to the $N=5$ case of \cite{KausReall}. 
We assume the electromagnetic field to be spherically symmetric and   purely  electric, yielding
\begin{align}\label{eme}
\star_{n}F = Q_{D}\dd\Omega_{D}
\end{align}
where $\star_{n}$ is the Hodge dual in $n$ dimensions, $F$ is the Faraday differential form, and $d\Omega_{D}$ is the volume form on a $D$ sphere. Given that for a $2-$form $\omega$ on a Lorentzian manifold the Hodge dual satisfies $\star_n\star_n\omega=-\omega$ 
the sign is $(-1)^{k(n-k)}s$ (where $s$ is the sign of our metric so (-1) and $k$ is the rank of the tensor)
we find that the electromagnetic field strength is given by 
\begin{align}
F =-\star_n\star_n F= -Q_{D}\star_n( \dd\Omega_D)=-Q_D\frac{S(k)A^2(z_0)}{R^D(z_0)}\dd t\wedge\dd r
\end{align}
We can use the definition of extrinsic curvature on our chosen metric and normal vector to write $K_{ab}=\left. -\frac{1}{2}\partial_z (g_{ab})\right|_{z=z_0}$. The junction conditions (\ref{junction}) then become \begin{align}
 \left. \frac{1}{2}\partial_z(g_{ab})\right|_{z=z_0}&=\frac{1}{\ell} h_{ab}+\frac{2\ell}{N-3}\left(\frac{3}{2(N-2)}\frac{Q_D^2 }{R^{2D}(z_0)}h_{ab}+\frac{Q_D^2A^{2}(z_0)}{R^{2D}(z_0)}\left(\delta_a^t\delta_b^t-S(k)^2\delta_a^r\delta_b^r\right)\right)
\end{align}
which reduce to  two independent constraints 
\begin{align}\label{israel-junction}
\frac{\alpha'(z_0/\ell)}{\alpha(z_0/\ell)}= 1 -\frac{2N-7}{(N-3)(N-2)} \frac{q_{D}^{2}}{\rho^{2D}(z_0/\ell)}\qquad \frac{\rho'(z_0/\ell)}{\rho(z_0/\ell)} = 1 + \frac{3}{(N-3)(N-2)}\frac{ q_{D}^{2}}{\rho^{2D}(z_0/\ell)}
\end{align}
in terms of $\alpha$ and $\rho$,
where $q_D=Q/\ell^{D-1}$. These conditions can be combined and rearranged to the form
\begin{align}\label{qrhocon}
3\frac{\alpha'(z_0/\ell)}{\alpha(z_0/\ell)}+(2D-1)\frac{\rho'(z_0/\ell)}{\rho(z_0/\ell)} = 2(D+1) \qquad q_{D} =\rho^{D}(z_{0})\sqrt{\frac{D}{2}\left(\frac{\rho'(z_0/\ell)}{\rho(z_0/\ell)}-\frac{\alpha'(z_0/\ell)}{\alpha(z_0/\ell)}\right)}
\end{align}
The Hamiltonian constraint can also be evaluated at $z=z_0$ to give 
\begin{align}\label{hamconrho}
\frac{7D-2}{D^2(D+1)}\frac{q_D^4}{\rho^{4D}(z_0/\ell)}+\frac{2(D-2)}{D }\frac{q_D^2}{\rho^{2D}(z_0/\ell)}+\frac{D(D-1)}{\rho^2(z_0/\ell)}=\frac{-2k}{\alpha^2(z_0/\ell)}
\end{align}
and,  for  $D\geq2$ each term on the left hand side is positive; hence   $k=-1$, eliminating the other choices. There may be interest in the D=1 case where $k$ does not have these restrictions from the Hamiltonian constraint.

\section{Solutions}

We can now restrict ourselves to $k=-1$ which affords us two exact solutions. We will first look at the properties of the two exact solutions and then explore the remainder of the parameter space. 

The two analytic solutions are a generalization of those found in \cite{KausReall}. For the first case we set $\beta = D+2$ or $\alpha(0)=1$ and find
\begin{align}\label{ana1}
\alpha(z/\ell) =\cosh(z/\ell) \qquad \rho(z/\ell)=\sinh(z/\ell)
 	\end{align}
However  the first condition in \eqref{qrhocon}   requires either
$z_0/\ell=\mathrm{arctanh}\left(\frac{1}{3}(2D-1)\right)$ or
$z_0/\ell=\mathrm{arctanh}(1)=\infty$.  The former solution is not real
for $D\geq 2$.  For the latter solution  both $\alpha=\infty$ and
$\rho=\infty$, forcing $q_D=0$. Under this \eqref{hamconrho} becomes
$D(D-1)=2$ for $D\geq2$, and so this analytic solution requires $D=2$.
As we will see in the following sections, for $D>2$ there are solutions
with $z_0/\ell=\infty$, which result from $\alpha(0)<1$; however they are
not analytic.

 

The second exact solution sets $\beta=1$ which results in $\alpha$ being constant
\begin{align}\label{ana2}
\alpha(z/\ell) = \frac{1}{\sqrt{D+2}} \qquad \rho(z/\ell) = \sqrt{\frac{D}{D+2}}\sinh\left(\sqrt{\frac{D+2}{D}}\frac{z}{\ell}\right).	
\end{align}
Using the Israel junction conditions, we find that we can express $q_{D},\rho_{0}$ exactly as
\begin{align}
z_{0} &=\ell \sqrt{\frac{D}{D+2}}\mathrm{arccoth}\left(\frac{2D+2}{2D-1}\sqrt{\frac{D}{D+2}}\right)\\
q_D &= \sqrt{\frac{D(D+1)}{2D-1}}\left((2D-1)\sqrt{\frac{D}{4D^2+11D-2}}\right)^D	
\end{align}

\begin{figure}
\begin{center}
\includegraphics[scale=0.8]{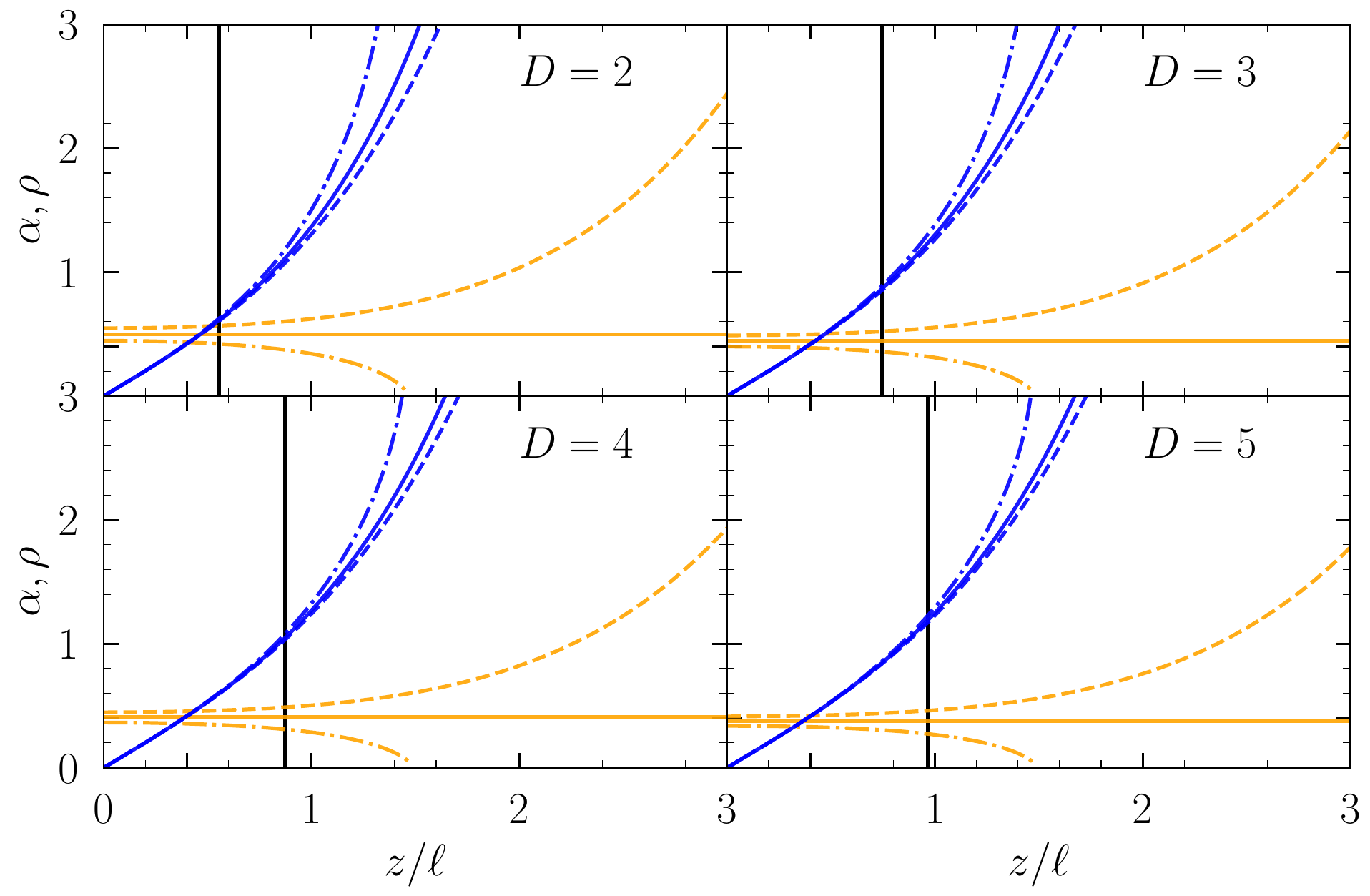}
\end{center}
\caption{Plots of the behaviour of $\rho(z/\ell)$ (blue)
and $\alpha(z/\ell)$ (orange). The dimension of
the hypersphere $D$ runs from $2$ to $5$ starting in the top left then
proceeding across and then down. $\beta$ takes the values of $1/2$ (dot
dashed), $1$ (solid) and $2$ (dashed). For $\beta<1$ we
place a vertical line at the location of the brane to demonstrate that
$z_0$ occurs before $\alpha=0$.}
\label{fig:alpha26}
\end{figure}

\begin{figure}
\begin{center}
 \includegraphics[scale=0.8]{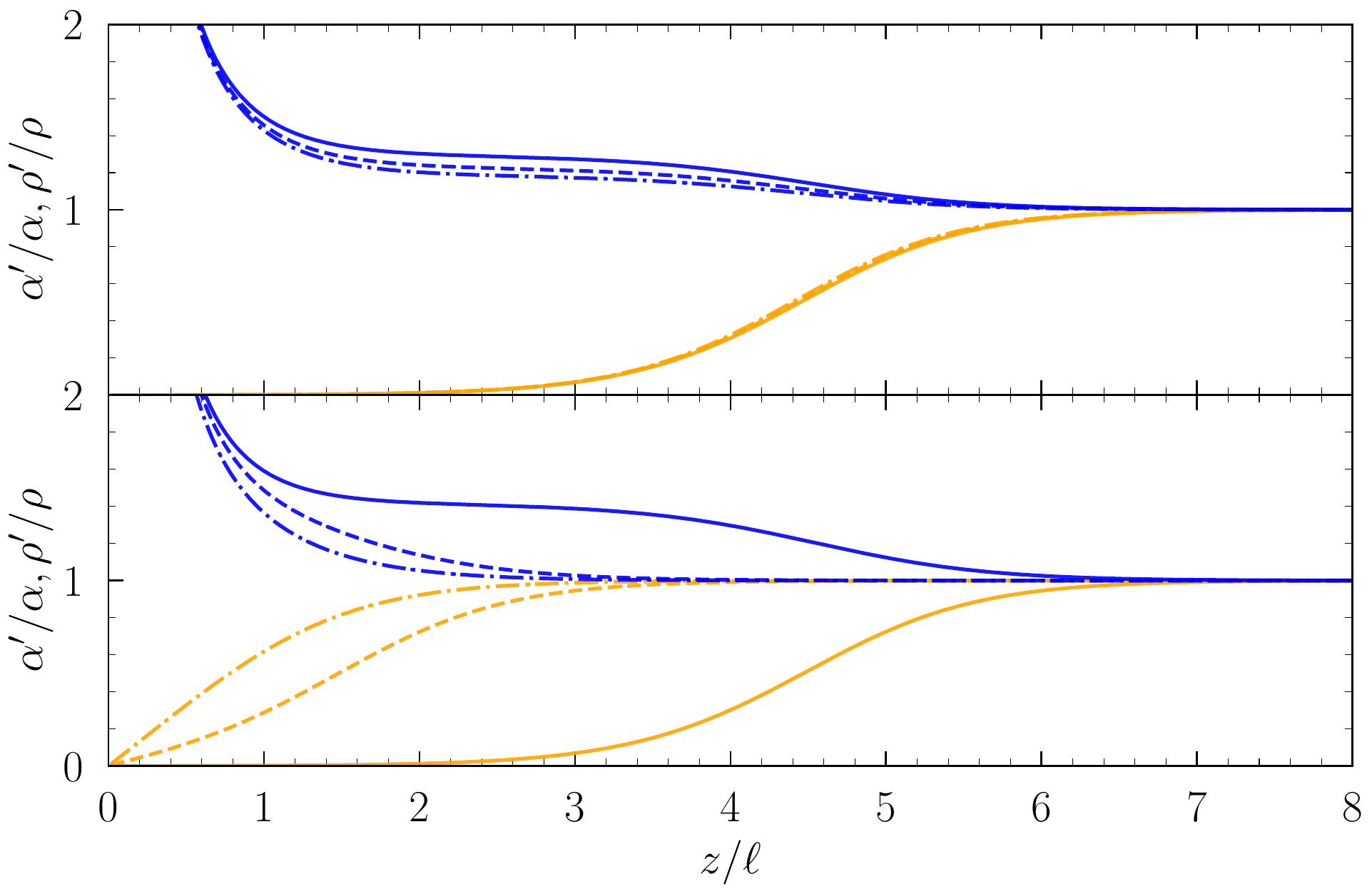}
\end{center}
\caption{$\rho'/\rho$ (blue) and $\alpha'/\alpha$ (orange) for various
$D=3,4,5$ (solid, dashed, dash dot) with fixed $\beta$ (top), and various
$\beta=1.001,1.2,2$ (solid, dashed, dash dot) with fixed $D$ (bottom).
This behaviour persists for all $\beta>1$ tested. We conclude
$\alpha(z/\ell),\rho(z/\ell)\propto e^{z/\ell}$ for large $z/\ell$.  }

\label{fig:rhoprime}
\end{figure}

For general values of $\beta$, we will rely on numeric techniques. As our
equations of motion are singular at $0$, we must rely on the expansions
(\ref{series1}) and (\ref{series2}) taken to
$\mathcal{O}(\left(\frac{z}{\ell}\right)^{22})$ which can be evaluated at
$z= 10^{-16}\ell $ to generate initial conditions. In general, we find
two kinds of behaviour surrounding the case of constant $\alpha$ which
can be seen in figure (\ref{fig:alpha26}). For $\beta>1$ we must have
$\alpha(0)>\sqrt{1/(D+2)}$, and we find that $\alpha$ and $\rho$ tend
towards being proportional to $e^{z/\ell}$ for large $z$, as illustrated
in figure (\ref{fig:rhoprime}). Conversely,  for $\beta<1$ we see new
behaviour where $\alpha$ monotonically decreases to $0$, for some finite
$z_1$, at which point $\rho$ also diverges. A calculation of the
Kretschmann scalar indicates that there is a curvature singularity at
$z_1$. This singularity would be naked if the brane, and therefore the
$\mathbb{Z}_2$ flip across it was not placed before it. In all our
simulations we find $z_0<z_1$ meaning the area of the solution is not
reached in the full RS model. 

\begin{figure}
\begin{center}
 \includegraphics[scale=0.8]{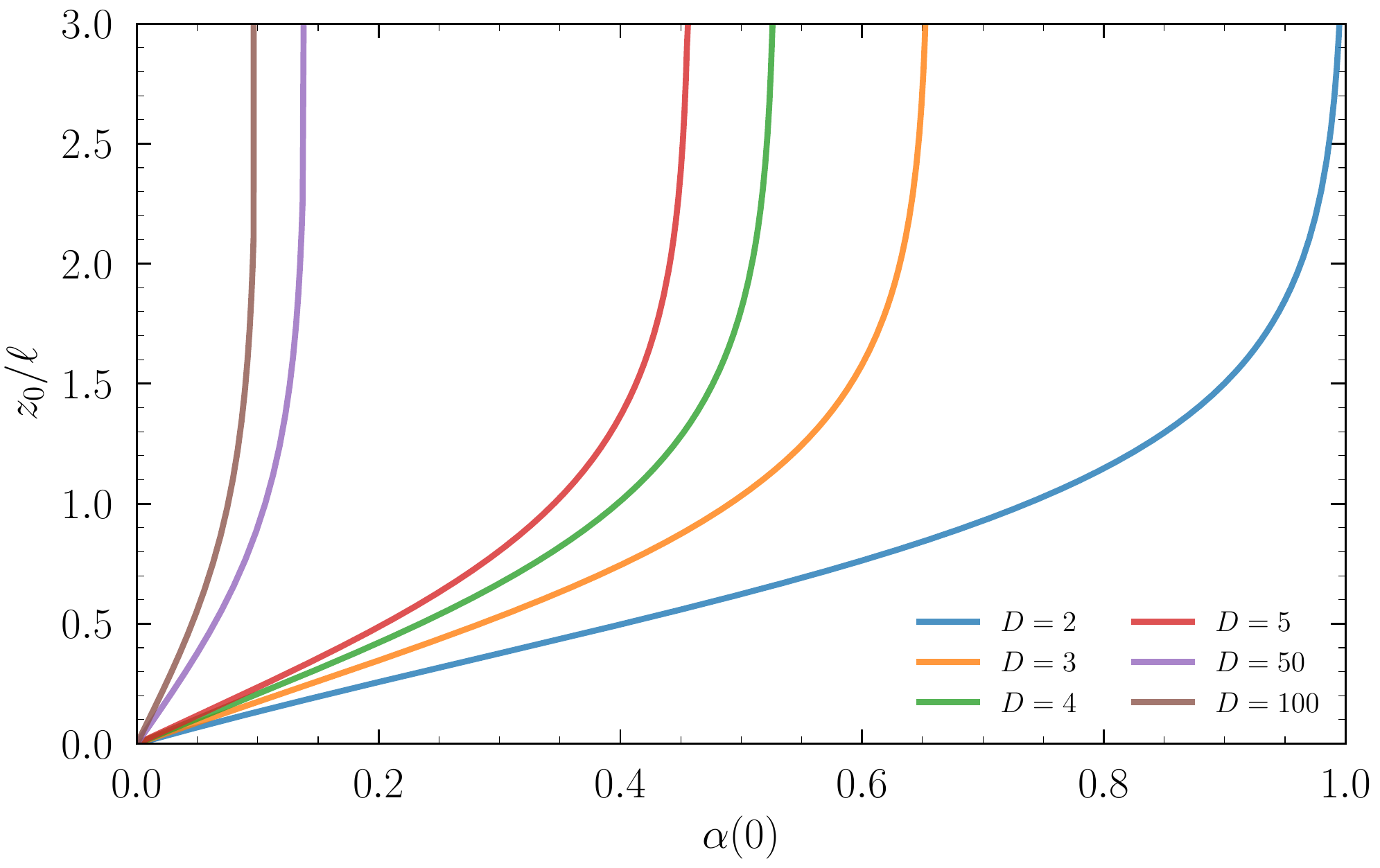}
\end{center}
\caption{The location of the brane $z_0/\ell$ of differing initial radii of the $AdS_2$ space for various spherical dimensions. }
\label{fig:alpha0}
\end{figure}

The equations of motion always have solutions and the junction conditions
eliminate additional regions parameter space present in these solutions.
For $D=2$ the limiting cases are the first analytic solutions
\eqref{ana1} and \eqref{ana2} described above \cite{KausReall}, and for
all $\alpha(0)<1$ solutions can be found. However for $D>2$ this is no
longer the case. We find that the value of the upper bound of $\alpha(0)$
is not easily characterized. In figure (\ref{fig:alpha0}) we plot
$z_0/\ell$ for various values of $\alpha(0)$ and we can see that $z_0$
seems to diverge for some value of $\alpha(0)$ that decreases with $D$,
see table (\ref{tb:abound}) for the value of the bounding $\alpha(0)$.
Below this bound, we find that solutions exist for all positive
$\alpha(0)$ with $z_0\propto\alpha(0)$ for small initial values. 

\begin{table}
\begin{center}
\begin{tabular}{|c|c|}
\hline 
Spherical Dimension ($D=N-3$) & Bound on $\alpha(0)$ \\ 
\hline 
2 & 1 \\ 
\hline 
3 & $\approx0.66$ \\ 
\hline 
4 & $\approx0.53$ \\ 
\hline 
5 & $\approx0.46$ \\ 
\hline 
6 & $\approx0.42$ \\ 
\hline
7 & $\approx0.38$ \\
\hline
8 & $\approx0.35$ \\
\hline
50 & $\approx0.14$ \\
\hline 
100 & $\approx0.10$\\
\hline
\end{tabular} 
\end{center}
\caption{The bounding value of $\alpha(0)$ of higher bulk dimension}\label{tb:abound}
\end{table}

\begin{figure}
\includegraphics[scale=0.8]{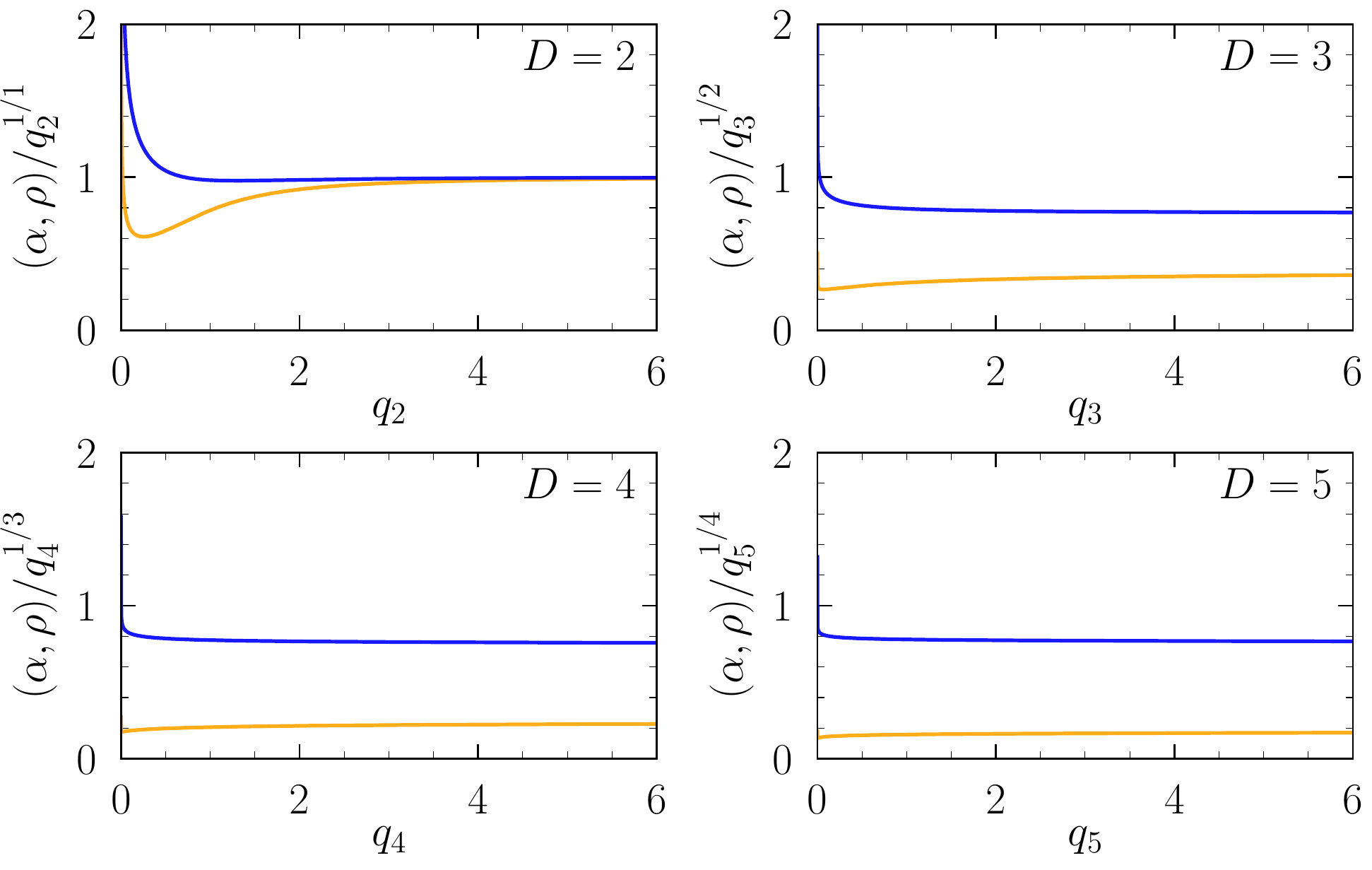}
\caption{The ratios between $\alpha(z_0/\ell)$  (orange) and
$\rho(z_0/\ell)$ (blue) to $q_D^{1/(D-1)}$.  The dimension of the
hypersphere $D$ runs from $2$ to $5$ starting in the top left then
proceeding across and then down.}
\label{fig:aqr}
\end{figure}

\begin{figure}
\begin{center}
\includegraphics[width=1.0\textwidth]{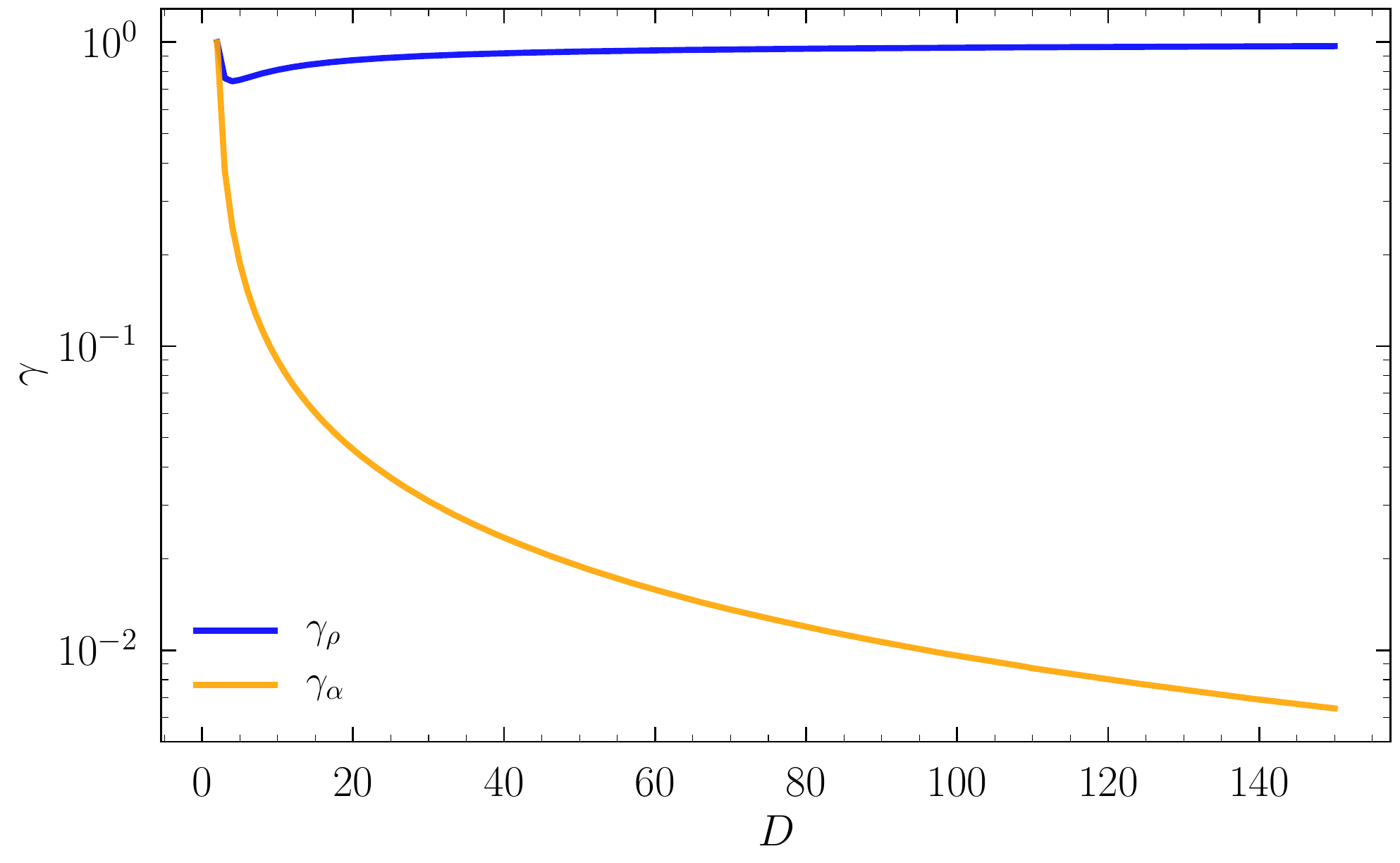}
\end{center}
\caption{Numerical ratio, $\gamma$, between $\alpha(z_0/\ell)$ (orange) and
$\rho(z_0/\ell)$ (blue) and $q_D^{1/(D-1)}$ for increasing dimension $D$.}
\label{fig:gammagamma}
\end{figure}

\newpage
We now switch to a parametrization in terms of the more physically
natural parameter of charge, $q_D$ to find the behaviour of
$\alpha(z_0),\rho(z_0)$ for large $z_{0}$. We can use the scaling of
$q_D$ with $\ell$ to guess that 

\begin{align}\label{asympradii}
\alpha(z_0)=\gamma_{(\alpha,D)} q_{D}^{1/D-1}\qquad
\rho(z_0)=\gamma_{(\rho,D)} q_{D}^{1/D-1} \qquad \text{for } q_D\gg 1
\end{align}
where $\gamma_{(\alpha,D)}$ and $\gamma_{(\rho,D)}$ are constants
dependent on dimension. Figure (\ref{fig:aqr}) demonstrates this
behaviour for the first few dimensions and figure (\ref{fig:gammagamma})
shows the behaviour of the numerically found constants.  We note that for
extremely small $q_D$ both $\alpha(z_0)/q^{1/(D-1)}$ and
$\rho(z_0)/q^{1/(D-1)}$ diverge, but the scale is incredibly small.

\section{Entropy of Small black holes}

Figure (\ref{fig:alpha0}) suggests that for small $z_0$, $z_{0}\propto
\alpha(0)$.  As such, by their expansions (\ref{series1}) and
(\ref{series2}) to first order, $\alpha(z_0)$ and $\rho(z_0)$ all scale
as $\alpha(0)$. From  the second relation of  (\ref{qrhocon}), we find
that $\rho(z_0)\propto q_{D}^{2/(2D-1)}$ in the regime of small $z_0$
where $\rho'(z_0)\approx 1$ and $\alpha'(z_0)=0$. Consequently, by
transitivity  $z_0, \alpha(z_0)$ and, $\alpha(0) \propto
q_{D}^{2/(2D-1)}$. Thus $\alpha(z_0)/q_D^{1/(D-1)}\propto
q_D^{-1/(2D-1)(D-1)}$,  which demonstrates the divergence of that ratio
for $q_D\rightarrow 0$. Identical scaling applies to $\rho(z_0)$ and
accounts for the similar behaviour of its ratio for small $q_D$. We then
find
$$
S_{D+3}\propto\int_0^{z_0}\dd z \rho^D(z/\ell)\approx\frac{\ell}{D+1}\rho^{D+1}(z_0)\propto\ell q_D^{2(D+1)/(2D-1)}
$$
unlike the expected scaling $\sim q_D^{{D}/{D-1}}$ from
\eqref{asympradii}. It is worth  noting  that unlike  the specific
case of $D=2$, for general $D$ we have  $\frac{D}{D-1}\neq \frac{2(D+1)}{2D-1}$
and as a result we find for small $z_0$ that our entropy scaling  changes
for $D>2$.



\section{Conclusion}

We have obtained extremally charged black hole solutions for
a higher dimensional Randall Sundrum models.  We find a dimensional
robustness to brane world black holes that is commensurate with previous 
$D=2$ results \cite{KausReall}.  While the equations of motion for the near
horizon geometry allow for the space to have a subspace that is any 2D
Lorentzian manifold, only the choice of $\text{AdS}_2$ allow for
satisfaction of the junction conditions.   Under these constraints we
find solutions in any dimension.

However we find that some properties of black holes for the $D=2$ case
are unique. For black holes that are large
compared to the AdS radius we find an identical scaling dependence of
entropy with respect to charge.  However in contrast to this, small black holes break the
scaling behaviour except for the specific case $D=2$. This change in
entropy scaling may be expected as the RS model can only recover
perturbative Newtonian gravity at which the scales are large compared to
$\ell$. For $D=2$ \cite{KausReall} it may just be a coincidence that such
matching holds for  smaller black holes. 

In the large $q_D$ regime we   also found that the constants of
proportionality for the scaling relations are no longer equal to one
another, with both becoming less than unity. As seen in
Figure ~(\ref{fig:gammagamma}) it appears the scaling for the AdS' subspace size
to $q_D$ falls off in the large dimension limit. The spherical subspace's
scaling distinquishes itself by initially decreasing from unity of the
initial $N=5$ case but slowly recovering becoming the larger contribution
to the size of the space. 
 
An exploration into how to better obtain the numerically found constant
of proportionality for entropy for given dimension or a formulation of
the dependence would also be advantageous. The divergence from the
standard entropy scaling relations leaves room for inquiry concerning on
if and how these new relations extend to non-static geometry systems. 


\section*{Acknowledgements}

This work was supported in part by the Natural Sciences and Engineering Research Council of Canada.

\appendix

\section*{Junction Condition}
\addcontentsline{toc}{chapter}{Appendix}

We generalize the junction conditions made for N=5 bulk in \cite{RSBH} to
bulk spacetimes of arbitrary dimension. We make use of Israel's technique
and break up our $N$ dimensional bulk into a family of $N-1$ dimensional
sub-manifolds described by the coordinates $x^a$, and a normal distance
from a particular surface $z$. To distinguish between tensors which lie
in the full space or only on the hypersurfaces, we will use Greek for the
former and Latin for the latter. Let the metric in the bulk take the form
$$\dd s^2=h_{ab}(x,z)\dd x^a\dd x^b+\dd z^2$$
where $h_{ab}$ lies in the tangent space of the sub-manifolds which
encapsulates the information in the intrinsic metric. We can also bring
$h$ into the full space to act as a projection metric to find the surface
parallel components of tensors in the tangent space of the bulk. To bring
$h$ up, let us define $n^\mu=\delta_z^\mu$ which describes the direction
normal to each surface. The projection tensor thus takes the form
\begin{equation}
\tilde{h}_{\mu\nu}=g_{\mu\nu}-n_\mu n_\nu.
\end{equation}

We will use the tilde to bring an element of the tangent space of the
$N-1$ dimensional structure into the tangent space of the $N$ dimensional
space via an inclusion map. The data encoded in $h$ could be employed to
find the intrinsic curvature on the surface, but we have more interest in
connecting the bulk's curvature to the $h$. In order to do this
connection, we need a means to measure the bending of the surface in the
larger space. This bending is measured using the extrinsic curvature
$$K_{\mu\nu}=\tilde{h}_\mu^\alpha\tilde{h}_\nu^\beta\nabla_\alpha n_\beta$$
which clearly is tangential to the hypersurface, and although subtle in
this form, it can be shown to be symmetric. There are a few results which
also follow from our choice of the Gauss Normal gauge. The first uses
that $n_\mu n^\mu=1$ for all coordinates which implies that
$0=\frac{1}{2}\nabla_\alpha(n_\mu n^\mu)=n^\mu\nabla_\alpha(n_\mu)$ .
Consequentially,

\begin{equation}
K_{\mu\nu}=\tilde{h}_\mu^\alpha\tilde{h}_\nu^\beta\nabla_\alpha n_\beta
=\tilde{h}_\mu^\alpha(\delta^\beta_\nu-n^\beta n_\nu)\nabla_\alpha n_\beta
=\tilde{h}_\mu^\alpha(\nabla_\alpha n_\nu- n_\nu n^\beta \nabla_\alpha n_\beta)
=\tilde{h}_\mu^\alpha\nabla_\alpha n_\nu.
\end{equation}
which allows us to not need the second projection. In fact, because
$n^\mu=\delta_z^\mu$ implying  $\partial_\mu n^\nu=0$ and
$g_{z\mu}=\delta^z_\mu$ one can conclude
$$
n^\mu\nabla_\mu n_\nu=n^\mu\left(\frac{1}{2}\partial_zg_{\mu\nu}\right)=\left(\frac{1}{2}\partial_zg_{z\nu}\right)=0.
$$
As a result of this property, we can further simplify the extrinsic curvature to require no projections

\begin{equation}\label{eq:noproj}
K_{\mu\nu}=\tilde{h}_\mu^\alpha\nabla_\alpha n_\nu
=(\delta^\alpha_\mu-n_\mu n^\alpha )\nabla_\alpha n_\nu
=\nabla_\mu n_\nu.
\end{equation}

Finally, we make note of two identities  
\begin{equation}
\nabla_\mu \tilde{h}_\alpha^\beta=-K_{\mu\alpha}n^\beta-{K_\mu}^\beta n_\alpha
\end{equation}
which follow from (\ref{eq:noproj}) and the Leibniz rule, and another for
the Lie derivative of the extrinsic curvature
\begin{equation}\label{eq:lie}
\mathcal{L}_n K_{\mu\nu}=n^\lambda\nabla_\lambda K_{\mu\nu}+K_{\lambda\nu}\nabla_\mu n^\lambda+K_{\mu\lambda}\nabla_\nu n^\lambda=n^\lambda\nabla_\lambda K_{\mu\nu}+2K_{\mu\lambda}{K_\nu}^\lambda.
\end{equation}
Using these tools, a well-known exercise yields 
 the Gauss-Codazzi relation
\begin{equation}
{\tilde{R}^{(N-1)\,\mu}}_{\;\;\;\qquad\lambda\alpha\beta}
=\left({K_{\alpha}}^{\mu} \tilde{h}^{\rho}_{\beta}-{K_{\beta}}^{\mu} \tilde{h}^{\rho}_{\alpha}\right)K_{\rho\lambda}
+\tilde{h}^{\mu}_{\mu'}\tilde{h}^{\lambda'}_{\lambda}\tilde{h}^{\alpha'}_\alpha \tilde{h}^{\beta'}_{\beta}{R^{\mu'}}_{\lambda'\alpha'\beta'}.
\end{equation}
and the relation
\begin{equation}
\mathcal{L}_n K_{\alpha\beta}={\tilde{R}^{(N-1)}}_{\qquad\alpha\beta}+2{K_\alpha}^\lambda K_{\beta\lambda}-8\pi G_N \tilde{h}_\alpha^{\alpha'}\tilde{h}_\beta^{\beta'}T_{\alpha'\beta'}+\left(\frac{N-1}{l^2}+\frac{8\pi G_N}{N-2}T\right)\tilde{h}_{\alpha\beta}.
\end{equation}
Thus, if we posit that there exists a infinitesimal surface of non zero
energy momentum, we can treat $T_{\alpha\beta}$ as the distribution
$\delta(z)T_{\alpha\beta}$. Integrating z from $(-\epsilon,\epsilon)$,
under the reasonable assumption of a finite discontinuity for all other
terms, we find in the limit as $\epsilon$ tends to 0

\begin{equation}\label{eq:IJC}
K_{\alpha\beta}(z=0^+)-K_{\alpha\beta}(z=0^-)=8\pi G_N\left(\frac{1}{N-2}T\tilde{h}_{\alpha\beta}-\tilde{h}_\alpha^{\alpha'}\tilde{h}_\beta^{\beta'}T_{\alpha'\beta'}\right)
\end{equation}
which constitute the Israel junction conditions.

\bibliography{HDRS-v14}
\bibliographystyle{ieeetr}

\end{document}